%% file: manuscript.tex
\documentclass[aps,prl,twocolumn,superscriptaddress,longbibliography]{revtex4-2}
\usepackage{textcomp}
\usepackage{amsmath}
\usepackage{cancel}
\usepackage{graphicx}

\makeatletter

\usepackage{xspace}
\usepackage{amsfonts}
\usepackage{color}
\definecolor{red}{rgb}{0.75,0,0}
\definecolor{blue}{rgb}{0,0,0.75}
\definecolor{green}{rgb}{0,0.5,0}
\definecolor{orange-red}{RGB}{255,69,0}

\newcommand{\eq}{\begin{equation}}
\newcommand{\eeq}{\end{equation}}
\newcommand{\aeq}{\begin{equation}\begin{aligned}}
\newcommand{\eaeq}{\end{aligned}\end{equation}}

\input{macros}

\makeatother

\begin{document}

\title{Nonlocal Sensing Drives Hybrid Phase Separation in Brownian Matter}
% A Perceptive Route to Hybrid Phase Separation
% Nonlocal perception-driven emergent order in Brownian matter

\author{Benchang Wu}
\thanks{These authors contributed equally}
\affiliation{Fujian Provincial Key Laboratory for Soft Functional Materials Research,
Research Institute for Biomimetics and Soft Matter, Department of
Physics, Xiamen University, Xiamen, Fujian 361005, China}

\author{Ziluo Zhang}
\thanks{These authors contributed equally}
\affiliation{Fujian Provincial Key Laboratory for Soft Functional Materials Research,
Research Institute for Biomimetics and Soft Matter, Department of
Physics, Xiamen University, Xiamen, Fujian 361005, China}

\author{Shutong Guo}
\thanks{These authors contributed equally}
\affiliation{School of Physics and Astronomy, Institute of Natural Sciences and
MOE-LSC, Shanghai Jiao Tong University, Shanghai 200240, China}

\author{Hepeng Zhang}
\thanks{Corresponding author: Hepeng\_zhang@sjtu.edu.cn}
\affiliation{School of Physics and Astronomy, Institute of Natural Sciences and
MOE-LSC, Shanghai Jiao Tong University, Shanghai 200240, China}
\author{Zhihong You}
\thanks{Corresponding author: zhyou@xmu.edu.cn}
\affiliation{Fujian Provincial Key Laboratory for Soft Functional Materials Research,
Research Institute for Biomimetics and Soft Matter, Department of
Physics, Xiamen University, Xiamen, Fujian 361005, China}

\date{\today}

%=====================================================
\begin{abstract}
Matter can organize not only through forces, but also through the information its constituents acquire from their surroundings. Here we use perceptive Brownian particles as a minimal model to isolate nonlocal sensing as an organizing principle for nonequilibrium matter. The particles undergo purely Brownian motion, with no mechanical interactions, self-propulsion, alignment, or auxiliary fields. Their only coupling is informational, through diffusivity regulated by density measured over a finite perception zone. Whereas local sensing, when unstable, produces conventional long-wavelength demixing, nonlocal perception restructures the instability spectrum, introducing finite-wavelength patterning and nonlinear bubbling instabilities. More fundamentally, it reshapes the ordering pathway by assembling a cascade of instabilities: macroscopic demixing creates dense domains, finite-wavelength modes pattern them internally, and nonlinear feedback hollows them into void bubbles. This produces hybrid phase separation, where a macroscopic dense phase coexists with a dilute background while retaining ordered internal microstructure, whose symmetry, anisotropy, and length scales are selected by the perception kernel. These results establish information acquisition as a constitutive principle of nonequilibrium matter, capable of governing both phase stability and the dynamical pathways through which order emerges.
\end{abstract}
\maketitle
%\tableofcontents

%====================================================
Information transfer offers a route by which nonequilibrium matter can couple its internal dynamics to its surroundings~\cite{Ramaswamy2010,Marchetti2013,Bechinger2016}. In many living and engineered systems, constituents do not merely respond to instantaneous forces or local thermodynamic variables. They sense their environment, process the acquired information through a perception rule, and use it to update how they move, fluctuate or interact~\cite{Couzin2005}. The material configuration then determines what is perceived, while the perceived information feeds back on the subsequent evolution of the configuration. This feedback suggests a form of self-organization that is not reducible to direct mechanical interactions or intrinsic motility alone. A central question is whether information acquisition can be isolated as a physical organizing principle in its own right, and whether the geometry by which matter perceives its environment can define new modes of phase ordering.

Active matter provides a natural arena for this question. Sustained nonequilibrium driving has been shown to generate collective phenomena such as motility-induced phase separation, flocking, clustering and active turbulence~\cite{Vicsek1995,TonerTu1995,Marchetti2013,Bechinger2016,Wensink2012,lopez2006macroscopic,zhou2024clustering,CatesTailleur2015}. When this driving is regulated by sensing or information exchange, the range of possible behaviours expands further~\cite{li2026informational,vansaders2026measurement}. Density-dependent motility and quorum sensing provide simple examples in which local population cues regulate activity and thereby couple sensing to aggregation or phase separation~\cite{TailleurCates2008,MillerBassler2001,WatersBassler2005,fischer2020quorum,Dinelli_2024,zhou2025visual,zhou2024clustering,zhou2023quorum,lefranc2025synthetic,rein2016collective}. More elaborate forms of communication arise in chemotactic, phoretic and field-mediated systems, where particles respond to signals produced by others, giving rise to long-ranged interactions, pattern formation and collective transport~\cite{KellerSegel1971,Golestanian2012,Saha2014,PohlStark2014,Liebchen2017,thewes2026phase}. Perception may also be geometrical, as in anisotropic or vision-based sensing, where the spatial region over which information is acquired encodes directional or nonlocal environmental cues~\cite{Ballerini2008,LiuColletive2025,zhou2025visual,lavergne2019group,negi2024collective,zhou2023quorum,saavedra2024self}. These advances show that sensing can strongly reshape active matter. 
Yet in most existing settings, perception remains tied to self-propulsion, alignment, chemical propulsion, external control or additional fields, making its independent role difficult to isolate. The key unresolved question is whether the way matter acquires information can itself generate a new route to self-organization, rather than merely renormalizing motility, modifying interactions or imposing external feedback.

\begin{figure*}[!t]
\centering
\includegraphics[width=\textwidth]{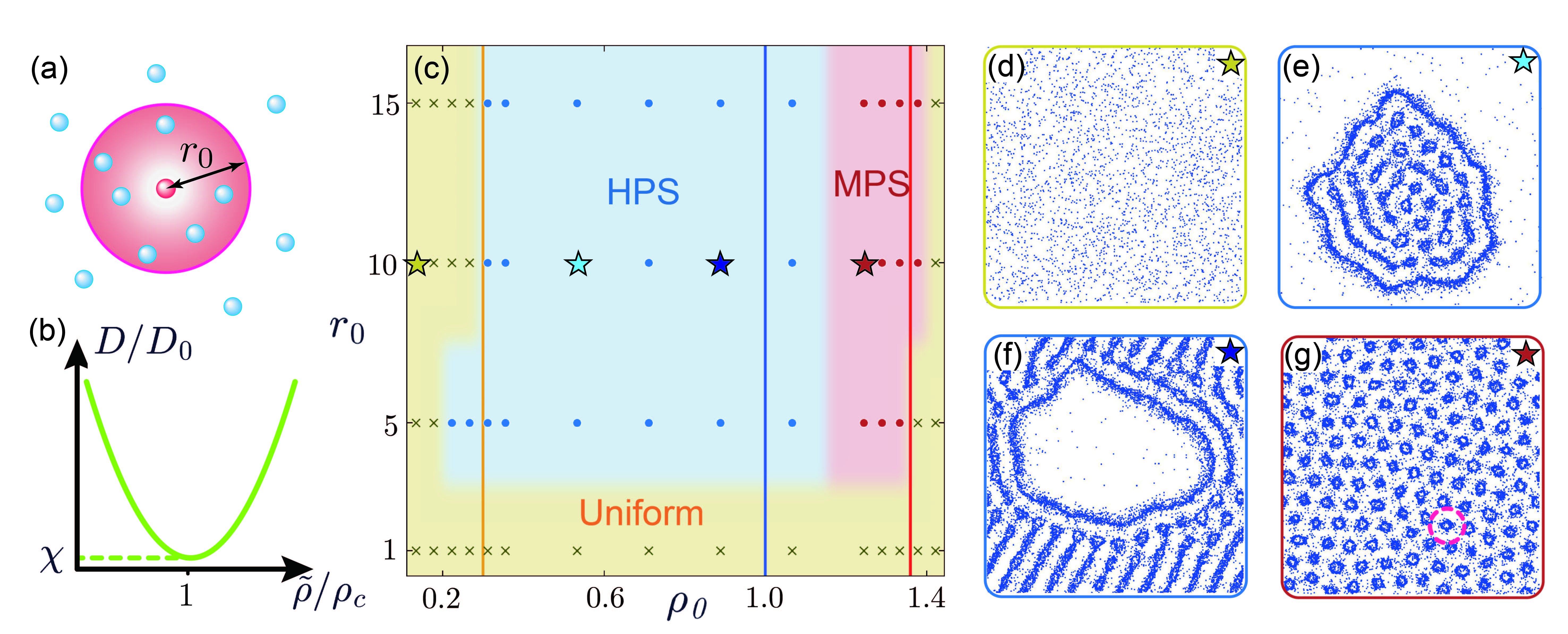}
\caption{\textbf{State diagram for circular perception.}
(a) Schematic of PBPs with a circular perception zone marked in red.
(b) Parabolic form of diffusivity $D$.
(c) State diagram at circular perception spanned by the radius $r_0$ and the mean density $\rho_0=N/L^2$. Solid lines denote state boundaries from the linear analysis of the continuum theory.
(d-g) Representative particle configurations at steady state in the (d) homogeneous, (e-f) HPS, (g) MPS states (red symbol shows the perception zone). The mean densities are: (d) $\rho_0=0.13$, (e) $\rho_0=0.53$, (f) $\rho_0=0.89$, and (g) $\rho_0=1.24$.}
\label{fig:PSPCirc}
\end{figure*}

Here we employ perceptive Brownian particles (PBPs), a minimal model in which otherwise passive Brownian particles regulate their diffusivity according to the density they perceive ~\cite{lopez2006macroscopic,zhou2024clustering,burger2013individual}. By removing self-propulsion, alignment and direct attractive interactions, the model exposes perception-mediated feedback as the essential source of nonequilibrium organization. We show that nonlocal perception does not simply shift the onset of phase separation. Instead, it reshapes the instability spectrum and couples long-wavelength demixing to finite-wavelength structuring, producing a perception-induced hybrid instability. The resulting state is hybrid phase separation: a macroscopic dense phase coexists with a dilute phase while developing ordered microstructures internally~\cite{Leibler1980,OhtaKawasaki1986,SeulAndelman1995}. The perception kernel acts as an information filter, so that its geometry selects the balance between macrophase separation and internal ordering, and thereby controls the hierarchy of length scales. As internal microstructure develops, nonlocal perception can further destabilize the microdomains through a nonlinear bubbling instability, generating ordered arrays of voids within the dense phase. These results establish that information acquisition reshapes nonequilibrium phase behavior by controlling both instability spectra and ordering pathways, placing the perception rule on equal footing with mechanical interactions.

%=================================================
\textit{Particle model.}
We consider a minimal model of $N$ overdamped Brownian particles in a two-dimensional box of size $L\times L$. The particles exert no direct mechanical forces and obey purely diffusive dynamics~\cite{lopez2006macroscopic,burger2013individual},
\begin{equation}
\label{eq:dtr}
\dot{\mathbf r}_i
=
\sqrt{2D(\tilde{\rho}_i)}\,\boldsymbol{\eta}_i(t),
\end{equation}
where $\boldsymbol{\eta}_i(t)$ is Gaussian white noise with
$\langle\eta_{i\alpha}(t)\eta_{j\beta}(t')\rangle
=\delta_{ij}\delta_{\alpha\beta}\delta(t-t')$.
The only coupling is informational: particle $i$ sets its diffusivity from the perceived density $\tilde{\rho}_i=n_i/S_i$, where $n_i$ is the number of particles, including itself, inside its perception zone of area $S_i$ (Fig.~\ref{fig:PSPCirc}a). The geometry of this zone specifies how density information is acquired and is the central control parameter. We use the parabolic response
$D(\tilde{\rho}_i)=D_0\left[\chi+\left(\tilde{\rho}_i/\rho_c-1\right)^2\right]$,
where $D_0$ sets the diffusivity scale, $\rho_c$ is the preferred perceived density, and $\chi D_0$ is a small bare diffusivity with $\chi\ll1$ (Fig.~\ref{fig:PSPCirc}b). Thus particles fluctuate least when $\tilde{\rho}_i=\rho_c$ and become more mobile when the perceived environment is either too dilute or too crowded, as in quorum-sensing systems, spacing-regulating swarms, and feedback-controlled colloids~\cite{MillerBassler2001,WatersBassler2005,fischer2020quorum,Rubenstein2014,fernandez2020feedback}. Since motility is regulated by density reconstructed through a perception zone rather than by the instantaneous local density, any organization arises from feedback between configuration, perception, and fluctuation regulation. We nondimensionalize lengths by $\ell=1/\sqrt{\rho_c}$ and times by $\tau=1/(D_0\rho_c)$. All results are reported in these units unless stated otherwise. Simulation details are given in Sec.~S1 of the SI.

%===============================================
\textit{Particle simulations.} 
To probe the effects of nonlocal perception, we first consider a circular perception region of radius $r_0$ (Fig.~\ref{fig:PSPCirc}a). Varying $r_0$ continuously tunes the range of information acquisition. Figure~\ref{fig:PSPCirc}c summarizes the steady states in the $(r_0,\rho_0)$ plane, where $\rho_0=N/L^2$. For sufficiently small $r_0$, perception is effectively local and the system remains visually homogeneous over the parameter range studied (Fig.~\ref{fig:PSPCirc}d and yellow region in Fig.~\ref{fig:PSPCirc}c). 
Once $r_0$ exceeds a critical value, nonlocal sensing destabilizes the homogeneous state and opens two ordered regimes inaccessible to local perception. At lower $\rho_0$, the system exhibits a multiscale ordering dynamics: dense clusters nucleate, hollow into bubble-bearing domains, and coarsen by coalescence and Ostwald ripening~\cite{LifshitzSlyozov1961,Wagner1961} (Fig.~\ref{fig:PSPCirc}e, Vid.~S1, and blue region in Fig.~\ref{fig:PSPCirc}c). The final state consists of a single internally patterned dense phase coexisting with a dilute background. This state is intrinsically multiscale: it emerges through large-scale phase separation while retaining finite-wavelength order within the dense phase, a signature of microphase separation~\cite{CahnHilliard1958,Bray1994,Leibler1980,OhtaKawasaki1986,SeulAndelman1995}. We therefore refer to this multiscale coexistence as \textit{hybrid phase separation} (HPS). As $\rho_0$ increases, the dense phase occupies a larger fraction of the system, whereas the internal bubble spacing remains nearly unchanged (Fig.~\ref{fig:PSPCirc}f, Vid.~S2). 
At still higher density, the dilute background disappears and the system crosses over to a microphase-separated state (MPS), where periodic bubble arrays span the entire system (Fig.~\ref{fig:PSPCirc}g, Vid.~S3).

\begin{figure}[t]
    \centering
    \includegraphics[width=0.48\textwidth]{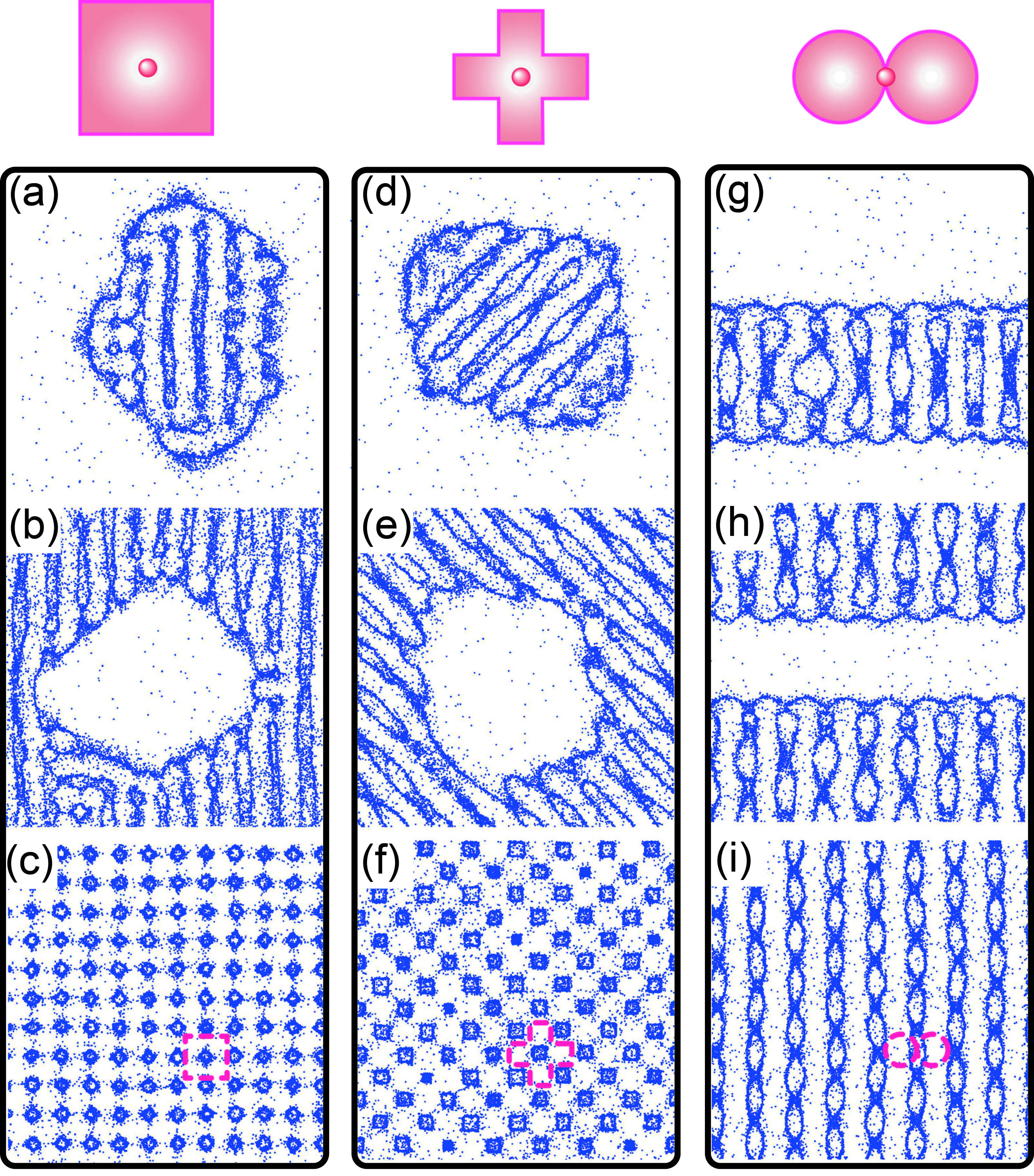}
    \caption{\textbf{Perception geometry determines the multiscale morphology of the emerging structures.}
    Columns show particle simulations with square, cross-shaped, and double-circle perception zones, respectively, while rows correspond to different mean densities.
    Red symbols in the bottom row indicate the corresponding perception zones. The mean densities are: (a,d,g) $\rho_0=0.53$, (b,e,h) $\rho_0=0.89$, (c) $\rho_0=1.51$, (f) $\rho_0=1.78$, (i) $\rho_0=1.24$.}    \label{fig:PSPGeom}
\end{figure}

Finite-range sensing opens a route to phase ordering, while sensing geometry selects the morphology of the ordered state. As shown in Fig.~\ref{fig:PSPGeom}, changing the perception kernel preserves the overall organization---HPS at lower density and MPS at higher density---but strongly reshapes the structures across scales. At the macroscale, anisotropic kernels stabilize anisotropic dense domains rather than the nearly circular droplets typical of conventional phase separation~\cite{Bray1994}. Square and cross-shaped perception both yield square-like droplets and cavities, but with interfaces aligned along diagonal and cardinal directions, respectively. Double-circle perception instead elongates dense domains along the sensing axis. This suggests that sensing anisotropy induces an effective anisotropic interfacial response that shapes the dense--dilute boundary. The internal microstructure is selected in the same way. Square and cross-shaped perception produce square-like microbubbles with different orientations, while double-circle perception yields elongated bubbles perpendicular to the sensing axis (Figs.~\ref{fig:PSPGeom}c,f,i). Bubble arrangements also reflect the kernel geometry, with each bubble occupying an exclusion neighborhood set by the perception zone. Thus the perception kernel acts as a morphological selector, imprinting its symmetry, anisotropy, and characteristic length scales onto the emergent multiscale morphology.

%=================================================
\textit{Continuum theory.}
To reveal the nonequilibrium mechanism underlying the emergent organization, we coarse-grain the particle dynamics using a Dean-type construction~\cite{Dean:1996} (Sec.~S2 of the SI). Retaining the deterministic contribution gives a conserved density dynamics,
\begin{equation}
    \partial_t \rho(\mathbf r,t)
    =
    \nabla^2 \mu_{\rm P}(\mathbf r,t)
    -
    \kappa\nabla^4\rho,
    \label{eq:continuum_dynamics}
\end{equation}
where the fourth-order term stabilizes short wavelengths~\cite{CahnHilliard1958,HohenbergHalperin1977}. Here, $\mu_{\rm P}(\mathbf r,t)=D\!\left[\bar{\rho}(\mathbf r,t)\right]\rho(\mathbf r,t)$ is the perception-mediated chemical potential, with $D(\bar{\rho})=D_0[\chi+(\bar{\rho}/\rho_c-1)^2]$ and $\bar{\rho}(\mathbf r,t)=\int d\mathbf r'\,K(\mathbf r-\mathbf r')\rho(\mathbf r',t)$.
$K$ is the normalized perception kernel, $\int d\mathbf r,K(\mathbf r)=1$. For circular perception, $K(\mathbf r)=\Theta(r_0-|\mathbf r|)/(\pi r_0^2)$. Unlike conventional chemical potentials tied to the local composition field~\cite{CahnHilliard1958,HohenbergHalperin1977,Wittkowski2014}, $\mu_{\rm P}$ is information-mediated: the density is first sampled over a finite perception range and only then converted into a local motility response. Perception therefore enters as a constitutive ingredient of Brownian matter, rather than merely renormalizing a local diffusivity. Numerical solutions of Eq.~\eqref{eq:continuum_dynamics} qualitatively reproduce the morphologies observed in the particle model (Sec. S3).

%=================================================
\begin{figure}[t]
    \centering
    \includegraphics[width=0.5\textwidth]{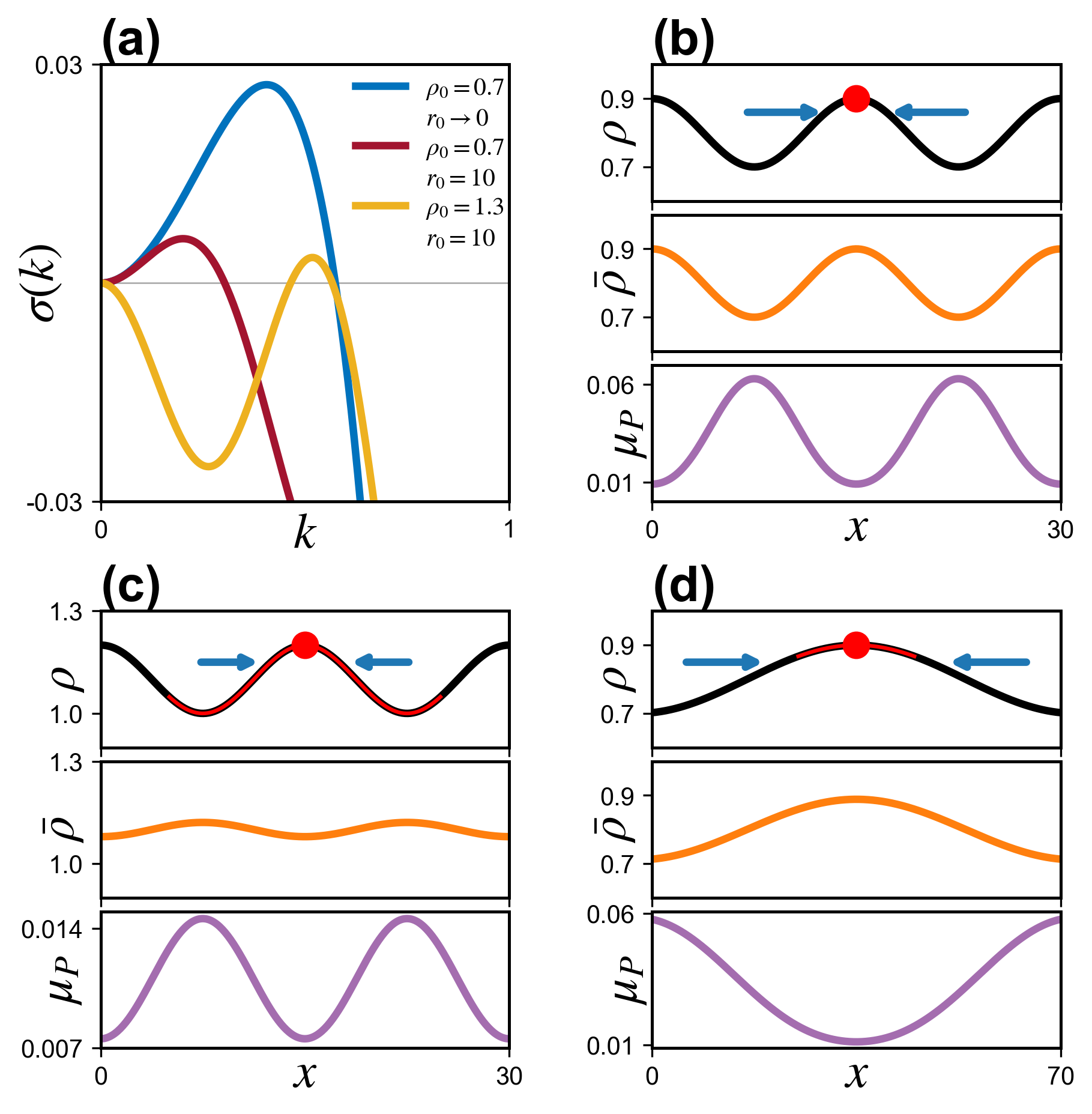}
    \caption{\textbf{Linear instability for circular perception.}
    (a) Dispersion relations $\sigma(k)$ for different perception radii $r_0$ and mean densities $\rho_0$.
    (b) In the local limit with $D'(\rho_0)<0$, the chemical potential $\mu_{\rm P}$ is out of phase with the density $\rho$, producing fluxes (blue arrows) that amplify the density fluctuation.
    (c) For finite-range perception with $D'(\rho_0)>0$, nonlocal filtering can make $\bar{\rho}$, and hence $\mu_{\rm P}$, out of phase with $\rho$, reversing the fluxes (blue arrows) so that material is driven into the density peak.
    (d) In the long-wavelength limit ($k\to0$) with $D'(\rho_0)<0$, the perceived density $\bar{\rho}$ closely follows the local density $\rho$, so finite-range perception does not change the long-wavelength nature of the instability.}
    \label{fig:LineStab}
\end{figure}

\textit{Nonlocal sensing reshapes phase ordering.} 
The continuum theory further shows that perception fundamentally reshapes nonequilibrium organization at the level of the instability spectrum. In conserved density dynamics, the nature of an instability is encoded in its dispersion relation: unstable modes near $k=0$ produce macrophase separation, while growth at finite $k$ produces intrinsic wavelength selection and yields pattern formation~\cite{CrossHohenberg1993,Brazovskii1975}. The striking feature of nonlocal sensing is that it does not merely shift the demixing threshold, but restructures the unstable spectrum and opens new pathways of phase ordering.

To see this, we linearize Eq.~\eqref{eq:continuum_dynamics} about a homogeneous state of density $\rho_0$ (Sec.~S4 of the SI). For perturbations $\delta\rho_{\mathbf k}\sim e^{\sigma(\mathbf k)t}$, the growth rate is
\begin{equation}
    \sigma(\mathbf k)
    =
    -k^2 \left[D(\rho_0)+
            \rho_0D'(\rho_0)\widehat K(\mathbf k)+\kappa k^2
        \right],
    \label{eq:sigmaCirc}
\end{equation}
with $k=|\mathbf k|$. The first term in brackets is the local diffusive response, while the second is the perception-mediated feedback, whose sign is controlled by $D'(\rho_0)\widehat K(\mathbf k)$. In the local limit, $K(\mathbf r)=\delta(\mathbf r)$ and $\widehat K(\mathbf k)=1$, so the only wavelength dependence comes from the stabilizing term $\kappa k^2$. Any instability is therefore attached to $k=0$ (blue line in Fig.~\ref{fig:LineStab}a) and requires
$D(\rho_0)+\rho_0D'(\rho_0)<0$. This criterion reflects the sign of the local chemical-potential response to a density perturbation. For local perception, $\mu_{\rm P}=D(\rho)\rho$, so $\delta\mu_{\rm P}=\left[D(\rho_0)+\rho_0D'(\rho_0)\right]\delta\rho$.
The homogeneous state is unstable only when a density increase lowers the local chemical potential. A density peak then becomes a chemical-potential minimum, and the current $\mathbf J=-\nabla\mu_{\rm P}$ drives material into the peak, amplifying the fluctuation (Fig.~\ref{fig:LineStab}b). Local perception thus supports only a Cahn--Hilliard-like long-wavelength demixing instability, analogous to the onset of MIPS~\cite{CahnHilliard1958,Bray1994,TailleurCates2008,CatesTailleur2015}. 

Finite-range perception fundamentally alters the linear-stability spectrum~\cite{TopazBertozziLewis2006,jewell2023patterning,thewes2026phase}. For circular perception of radius $r_0$, the Fourier kernel is $\widehat K(\mathbf k)=2J_1(kr_0)/(kr_0)$, where $J_1$ is the first-order Bessel function. Because $\widehat K(\mathbf k)$ is nonmonotonic and sign-changing, the dispersion relation becomes wavelength selective: modes unstable in the local theory can be suppressed, while locally stable modes can become unstable at perception-selected wavelengths. This is clearest for $\rho_0>\rho_c$, where $D'(\rho_0)>0$. In such case, the homogeneous state is always stable at local sensing, since the chemical potential $\mu_{P}=D(\rho)\rho$ increases monotonically with the local density. With finite-range perception, however, negative lobes of $\widehat K(\mathbf k)$ make the feedback term $\rho_0D'(\rho_0)\widehat K(\mathbf k)$ destabilizing at finite $k$. When this contribution overcomes $D(\rho_0)+\kappa k^2$, $\sigma(k)$ develops a finite-wavelength unstable band (orange line in Fig.~\ref{fig:LineStab}a). Intuitively, nonlocal filtering can make the perceived-density modulation, and hence $\mu_{\rm P}$, out of phase with the actual density field, reversing the diffusive flux and amplifying density peaks (Fig.~\ref{fig:LineStab}c). This mechanism accounts for the MPS observed at large $\rho_0$ in Figs.~\ref{fig:PSPCirc} and \ref{fig:PSPGeom}~\cite{Leibler1980,OhtaKawasaki1986,SeulAndelman1995}. Consistently, linear stability predicts both the ordering threshold (red line in Fig.~\ref{fig:PSPCirc}c) and the characteristic structures: the dominant peaks in the simulated structure factors coincide with the fastest-growing modes of $\sigma(k)$ (Sec.~S4). Nonlocal perception therefore destabilizes density modes that are stable in the local limit, turning information range into a wavelength-selection mechanism.

For $\rho_0<\rho_c$, where $D'(\rho_0)<0$, finite-range perception modifies the growth rates but not the long-wavelength character of the primary instability, since $\widehat K(k)\to1$ as $k\to0$ (red line in Fig.~\ref{fig:LineStab}a and Fig.~\ref{fig:LineStab}d). Thus, when unstable, the homogeneous state first follows the conventional Cahn--Hilliard route: long-wavelength fluctuations grow and separate the system into dense and dilute regions~\cite{CahnHilliard1958,Bray1994}. As demixing proceeds, however, the density inside the dense domains $\rho_d$ can rise to $\rho_d>\rho_c$, where $D'(\rho_d)>0$. The sign-changing perception kernel can then destabilize finite-$k$ modes within the dense phase, inducing a secondary pattern-forming instability, as illustrated by the finite-$k$ unstable branch (orange line in Fig.~\ref{fig:LineStab}a). The dense domains thereby acquire internal order while macroscopic phase separation persists. This hierarchical sequence---large-scale demixing followed by internal wavelength selection---produces HPS, showing that nonlocal sensing restructures not only the instability threshold but also the route of phase ordering.

%=================================================
\begin{figure}[t]
    \centering
    \includegraphics[width=0.48\textwidth]{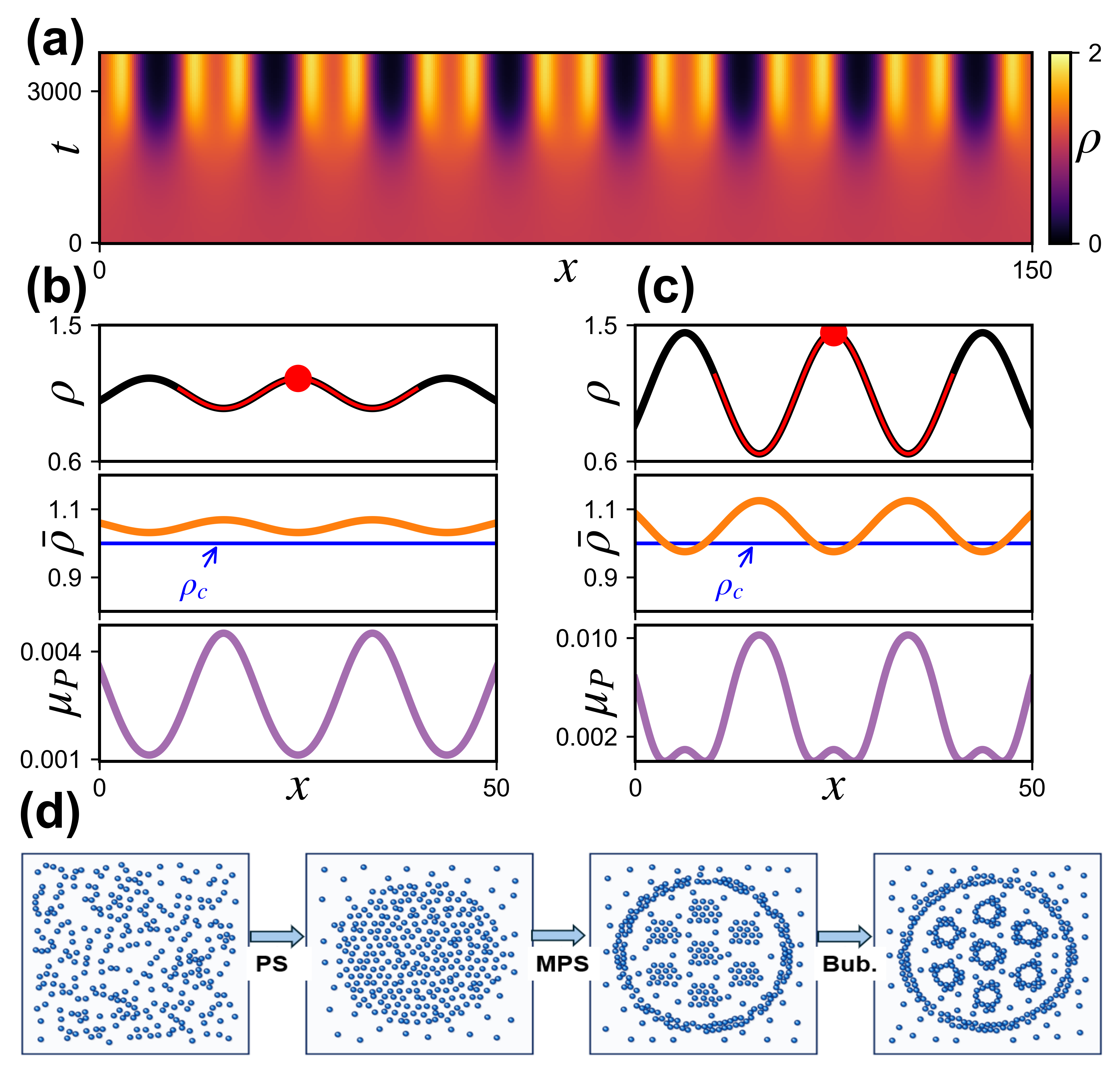}
    \caption{\textbf{Nonlinear mechanism of bubbling.}
    (a) Spatiotemporal evolution of the one-dimensional density field $\rho(x,t)$, showing the growth of a finite-wavelength mode and its nonlinear transformation into a bubbling state. See Sec.~S3C of the SI for simulation details.
    (b,c) Chemical-potential response generated by representative single-mode profiles $\rho(x)\simeq \rho_0+A\cos(k_*x)$.
    (b) For small amplitude $A$, the perceived density at the peak remains above $\rho_c$, and the perception-mediated chemical potential reinforces the density modulation.
    (c) For large $A$, nonlocal filtering lowers the perceived density at the peak below $\rho_c$, reversing the local curvature of $\mu_{\rm P}$ and driving material outward from the peak center, thereby initiating hollowing.
    (d) Schematic illustration of the cascade of instabilities leading to HPS.
}
    \label{fig:bubble}
\end{figure}

% ===================================================
\textit{Nonlocal sensing induced bubbling.} 
While linear stability explains the onset of HPS and MPS, it does not account for the robust emergence of bubbles across sensing geometries. These bubble-based microstructures differ from conventional microphase-separated states, whose elementary units are typically compact dense domains~\cite{Leibler1980,OhtaKawasaki1986,SeulAndelman1995}. To examine the onset of bubbling more directly, we simulate a one-dimensional version of Eq.~\eqref{eq:continuum_dynamics} in the MPS regime (Fig.~\ref{fig:bubble}a and Sec.~S3C). Starting from a nearly uniform state, the fastest-growing mode first grows as predicted by Eq.~\eqref{eq:sigmaCirc}. At later times, the density maxima develop central indentations that deepen into a one-dimensional bubbling state. The mechanism follows from how finite-range perception filters a growing density peak. For an early profile $\rho(x)=\rho_0+A\cos(k_*x)$, where $k_*$ is the most unstable mode, the perceived density is $\bar{\rho}(x)=\rho_0+A\widehat K(k_*)\cos(k_*x)$. In the MPS regime considered here, where $D'(\rho_0)>0$, the finite-$k_*$ instability requires $\widehat K(k_*)<0$, the peak center perceives an increasingly depleted environment as $A$ grows (Fig.~\ref{fig:bubble}b). At small amplitude, the perceived density at the peak remains above $\rho_c$, so the peak center is a chemical-potential minimum and continues to accumulate mass. When $A>(\rho_0-\rho_c)/|\widehat K(k_*)|$, the perceived density crosses $\rho_c$, reversing the local curvature of $\mu_{\rm P}$ (Fig.~\ref{fig:bubble}c and Sec.~S5). The density peak then becomes a chemical-potential maximum, driving a conserved outward flux toward the shoulders. The compact density maximum consequently flattens, hollows, and becomes a bubble. Thus nonlocal perception converts finite-wavelength density peaks into void-bearing domains through a nonlinear inversion of the perception-mediated chemical-potential landscape.

%=================================================
\textit{Discussion.}
We have used perceptive Brownian particles as a minimal model to isolate how finite-range sensing organizes nonequilibrium Brownian matter. Previous works on related Brownian models showed that density-dependent mobility can produce aggregation, demixing, or pattern formation~\cite{lopez2006macroscopic,zhou2024clustering,thewes2026phase}. Here we show that finite-range perception plays a more fundamental role: it not only reshapes the instability spectrum, but also reorganizes the route to order. In the system studied here, nonlocal sensing produces HPS at lower densities, MPS at higher densities, and robust bubbling within the ordered microstructures. In particular, the HPS state arises through a cascade of perception-mediated instabilities (Fig.~\ref{fig:bubble}d): long-wavelength demixing first forms macroscopic dense domains, finite-wavelength modes then pattern those domains internally, and nonlinear feedback finally hollows the microdomains into void bubbles. The resulting multiscale morphology is further sculpted by the perception kernel, whose geometry selects the symmetry, anisotropy, and characteristic length scales of the ordered state.

This picture is distinct from previous routes to phase ordering, which typically rely on competing interactions, self-propulsion, chemical fields, or explicit interparticle forces~\cite{Leibler1980,OhtaKawasaki1986,SeulAndelman1995,Wittkowski2014,Nardini2017,TjhungNardiniCates2018}. Here the particles are otherwise Brownian, and the nonlocal kernel is not mechanical but an information-acquisition rule that filters the density field before it is converted into motility. Sensing therefore does not merely renormalize motility or shift instability thresholds. It enters as a constitutive ingredient of nonequilibrium matter, determining both phase stability and the sequence of instabilities through which ordered states are dynamically assembled. More broadly, our results suggest that nonequilibrium phases can be shaped not only by forces, activity, and conservation laws, but also by how constituents measure their surroundings. This points toward a broader statistical mechanics of adaptive matter, in which sensing kernels and information-processing rules enter as constitutive laws that shape phases, mechanics, interfaces, fluctuations, and defects.

% ========================================================

\bibliographystyle{apsrev4-2}
\bibliography{QS_Brownian}

% ========================================================
\section*{Acknowledgements}
ZY was supported by the National Key Research and Development Program of China (No. 2023YFA1407500), the National Natural Science Foundation of China No. 12374219, the 111 project (B16029). HPZ was supported by the National Natural Science Foundation of China (No. 12225410, No. 12074243) and by the China Manned Space Engineering Program (Project number: KJZ-YY-NLT0502).

\section*{Conflicts of interest}
There are no conflicts of interest.

% ========================================================

\end{document}

\title{Supplemental Material for ``Nonlocal Sensing Drives Hybrid Phase Separation in Brownian Matter''}

\author{Benchang Wu}
\thanks{These authors contributed equally}
\affiliation{Fujian Provincial Key Laboratory for Soft Functional Materials Research,
Research Institute for Biomimetics and Soft Matter, Department of
Physics, Xiamen University, Xiamen, Fujian 361005, China}

\author{Ziluo Zhang}
\thanks{These authors contributed equally}
\affiliation{Fujian Provincial Key Laboratory for Soft Functional Materials Research,
Research Institute for Biomimetics and Soft Matter, Department of
Physics, Xiamen University, Xiamen, Fujian 361005, China}

\author{Shutong Guo}
\thanks{These authors contributed equally}
\affiliation{School of Physics and Astronomy, Institute of Natural Sciences and
MOE-LSC, Shanghai Jiao Tong University, Shanghai 200240, China}

\author{Hepeng Zhang}
\thanks{Corresponding author: Hepeng\_zhang@sjtu.edu.cn}
\affiliation{School of Physics and Astronomy, Institute of Natural Sciences and
MOE-LSC, Shanghai Jiao Tong University, Shanghai 200240, China}
\author{Zhihong You}
\thanks{Corresponding author: zhyou@xmu.edu.cn}
\affiliation{Fujian Provincial Key Laboratory for Soft Functional Materials Research,
Research Institute for Biomimetics and Soft Matter, Department of
Physics, Xiamen University, Xiamen, Fujian 361005, China}

\date{\today}

\maketitle

\newcommand{\rhoc}{\rho_c}

% ===============================================
\section{Simulation protocol for the particle model}

We simulate the particle dynamics by integrating Eq.~(1) of the main text with the Euler--Maruyama scheme~\cite{kloeden1992numerical} using a time step $\delta t=0.1$. At each step, the perceived density $\tilde{\rho}_i(t)$ is evaluated from the particle configuration before the update. The corresponding diffusivity $D[\tilde{\rho}_i(t)]$ is then assigned, and the particle position is advanced according to
\begin{equation}
    \mathbf r_i(t+\delta t)
    =
    \mathbf r_i(t)
    +
    \sqrt{2D[\tilde{\rho}_i(t)]\delta t}\,
    \boldsymbol{\zeta}_i ,
\end{equation}
where $\boldsymbol{\zeta}_i$ is a vector of independent Gaussian random variables with zero mean and unit variance.

This update corresponds to the It\^o convention and reflects the causal structure of perception-regulated Brownian motion: a particle first senses its environment, uses the acquired information to set its diffusivity, and then undergoes a stochastic displacement. Equivalently, the noise amplitude during $[t,t+\delta t]$ is determined only by information available at time $t$.

For each particle, the perceived density is $\tilde{\rho}_i=n_i/S_i$, where $n_i$ is the number of particles inside the prescribed perception zone (including itself) and $S_i$ is its area. The perception zone is centered on the particle and translated with it. Unless otherwise stated, simulations are performed in the dimensionless units defined in the main text, with length measured in $1/\sqrt{\rho_c}$ and time in $1/(D_0\rho_c)$. We use a square domain of size $L\times L$ with periodic boundary conditions and initialize the system from nearly uniform random particle positions. The parameters used for each figure are listed in Table~\ref{tab:simulation_parameters}.

\begin{table*}[h]
\caption{\label{tab:simulation_parameters}
Simulation parameters and sensing-kernel geometries.  Each sensing kernel is centered at the
origin and normalized by its support area. All values are measured in dimensionless
units.}
\begin{ruledtabular}
\begin{tabular}{lccc}
Quantity & Symbol & Value & Definition \\
\hline
System size & $L$ & $150$ & $L_x=L_y=L$ \\
Time step & $\delta t$ & $0.1$ & integration time step \\
Particle basic diffusion coefficient & $\chi$ & $2.0\times10^{-3}$ & basic bare diffusive coefficient \\
\hline
Disk radius & $r_0$ & $10$ & radius of the circular sensing domain \\
Square width & $a$ & $20$ & side length of the square sensing domain \\
Cross length & $l$ & $30$ & length of each rectangle of the cross \\
Cross width & $w$ & $10$ & width of each rectangle of the cross \\
Double-circle radius & $r_0'$ & $7.5$ & radius of each circle \\
\end{tabular}
\end{ruledtabular}
\end{table*}

% ========================================================
\section{Derivation of the continuum equations}

Here we derive the stochastic density equation associated with the particle dynamics. We start from the Langevin equation under the Itô convention
\begin{equation}
\dot{\mathbf r}_i
=
\sqrt{2D_i}\boldsymbol{\eta}_i(t),
\qquad
D_i \equiv D(\tilde{\rho}_i)=D_0[\chi+(\tilde{\rho}_i/\rho_c-1)^2],
\end{equation}
where $\boldsymbol{\eta}_i(t)$ is Gaussian white noise satisfying
$\langle \eta_{i\alpha}(t)\eta_{j\beta}(t')\rangle
=\delta_{ij}\delta_{\alpha\beta}\delta(t-t')$.
The diffusivity $D_i$ depends on the particle configuration through the perceived density $\tilde{\rho}_i$. 

Following Dean's construction~\cite{Dean:1996}, we define the microscopic density
\begin{equation}
\rho(\mathbf r,t)
=
\sum_{i=1}^N
\delta[\mathbf r-\mathbf r_i(t)] .
\end{equation}
For an arbitrary smooth test function $f(\mathbf r)$,
\begin{equation}
f[\mathbf r_i(t)]
=
\int d\mathbf r
\delta[\mathbf r-\mathbf r_i(t)]f(\mathbf r).
\end{equation}
Applying Ito's lemma gives
\begin{equation}
d f[\mathbf r_i(t)]
=
\sqrt{2D_i}\nabla f[\mathbf r_i(t)]\cdot d\mathbf W_i
+
D_i\nabla^2 f[\mathbf r_i(t)]dt ,
\end{equation}
where $d\mathbf W_i$ is a Wiener increment. Rewriting the right-hand side in terms of the microscopic density and integrating by parts, one obtains the exact stochastic equation
\begin{equation}
\partial_t \rho(\mathbf r,t)
=
\sum_{i=1}^N
\nabla^2
\left[
D_i \delta(\mathbf r-\mathbf r_i)
\right]
-
\nabla\cdot
\sum_{i=1}^N
\left[
\sqrt{2D_i}
\delta(\mathbf r-\mathbf r_i)
\boldsymbol{\eta}_i(t)
\right].
\end{equation}

For the perceptive Brownian particles considered here, the particle diffusivity can be written as a field evaluated at the particle position,
\begin{equation}
D_i
=
D[\bar{\rho}(\mathbf r_i,t)] =D_0[(\bar{\rho}/\rhoc-1)^2+\chi],
\end{equation}
where the perceived density is the nonlocal average
\begin{equation}
\bar{\rho}(\mathbf r,t)
=
\int d\mathbf r'
K(\mathbf r-\mathbf r')\rho(\mathbf r',t).
\end{equation}
The perception kernel is normalized as $\int d\mathbf r,K(\mathbf r)=1$ and is set by the geometry of the perception zone. For example, for a perception zone of area $S$, $K=1/S$ inside the zone and $K=0$ outside. Using the sifting property of the delta function, the deterministic part becomes
\begin{equation}
\sum_{i=1}^N
\nabla^2
\left[
D_i\delta(\mathbf r-\mathbf r_i)
\right]
=
\nabla^2
\left[
D(\bar{\rho})\rho
\right].
\end{equation}

The noise term can be written in Dean form by introducing a Gaussian vector white-noise field $\boldsymbol{\xi}(\mathbf r,t)$ with
\begin{equation}
\left\langle
\xi_\alpha(\mathbf r,t)
\xi_\beta(\mathbf r',t')
\right\rangle
=
\delta_{\alpha\beta}
\delta(\mathbf r-\mathbf r')
\delta(t-t') .
\end{equation}
Matching the noise correlations gives
\begin{equation}
-\nabla\cdot
\sum_{i=1}^N
\left[
\sqrt{2D_i}
\delta(\mathbf r-\mathbf r_i)
\boldsymbol{\eta}_i(t)
\right]
\equiv
\nabla\cdot
\left[
\sqrt{2D(\bar{\rho})\rho}
\boldsymbol{\xi}(\mathbf r,t)
\right],
\end{equation}
where the overall sign has been absorbed into the definition of $\boldsymbol{\xi}$. The stochastic density equation is therefore
\begin{equation}
\partial_t \rho(\mathbf r,t)
=
\nabla^2
\left[
D(\bar{\rho})\rho
\right]
+
\nabla\cdot
\left[
\sqrt{2D(\bar{\rho})\rho}
\boldsymbol{\xi}(\mathbf r,t)
\right].
\label{eq:deans_nonlocal}
\end{equation}
Equation~\eqref{eq:deans_nonlocal} is the Dean equation for Brownian particles whose diffusivity is regulated by a nonlocal perceived density.

In the main text, we focus on the deterministic mean-field dynamics that controls the instability spectrum and morphology selection. We therefore neglect the multiplicative noise in Eq.~\eqref{eq:deans_nonlocal} and add a stabilizing fourth-order gradient term to regularize short wavelengths. This gives
\begin{equation}
\partial_t \rho(\mathbf r,t)
=
\nabla^2
\left[
D(\bar{\rho})\rho
\right]
-
\kappa\nabla^4\rho(\mathbf r,t),
\label{eq:continuum_main}
\end{equation}
which is the continuum equation analyzed in the main text. Equivalently, Eq.~\eqref{eq:continuum_main} can be written as
\begin{equation}
\partial_t\rho
=
\nabla^2\mu_{\rm P}
-
\kappa\nabla^4\rho,
\qquad
\mu_{\rm P}=D(\bar{\rho})\rho .
\end{equation}
Here $\mu_{\rm P}$ is a perception-mediated chemical potential: unlike a local chemical potential, it depends on the density field sampled over a finite perception range before being converted into a local transport response.

\section{Simulations of field theory}
\label{sec:SimulationsOfFiledTheory}
% -----------------------------------------------------
\subsection{Simulation protocol}

To simulate the continuum theory, we solve Eq.~(2) of the main text on a two-dimensional square grid using a finite-difference scheme~\cite{burden2015numerical}. Unless otherwise stated, the grid spacing is $h=1$,  the time step is $\Delta t=0.5$ in the dimensionless units of the particle model and the stabilizing hyperdiffusion coefficient is $\kappa=0.001$. At each time step, the perceived density is computed from the nonlocal convolution $\bar\rho(\mathbf r,t)=\int d\mathbf r'\,K(\mathbf r-\mathbf r')\rho(\mathbf r',t)$ via FFT with the periodic boundary conditions, and using the same perception kernel as in the particle simulations. The perception-mediated chemical potential $\mu_{\rm P}=D(\bar\rho)\rho$ is then evaluated locally, and the density field is advanced according to Eq.~(2) of the main text. The total mass is conserved up to numerical accuracy throughout the simulations. Unless otherwise stated, all other parameters are the same as those used in the particle model and listed in Table~\ref{tab:simulation_parameters}.

% -------------------------------------------------------
\subsection{Simulation results}
\begin{figure}
    \centering
    \includegraphics[width=0.9\linewidth]{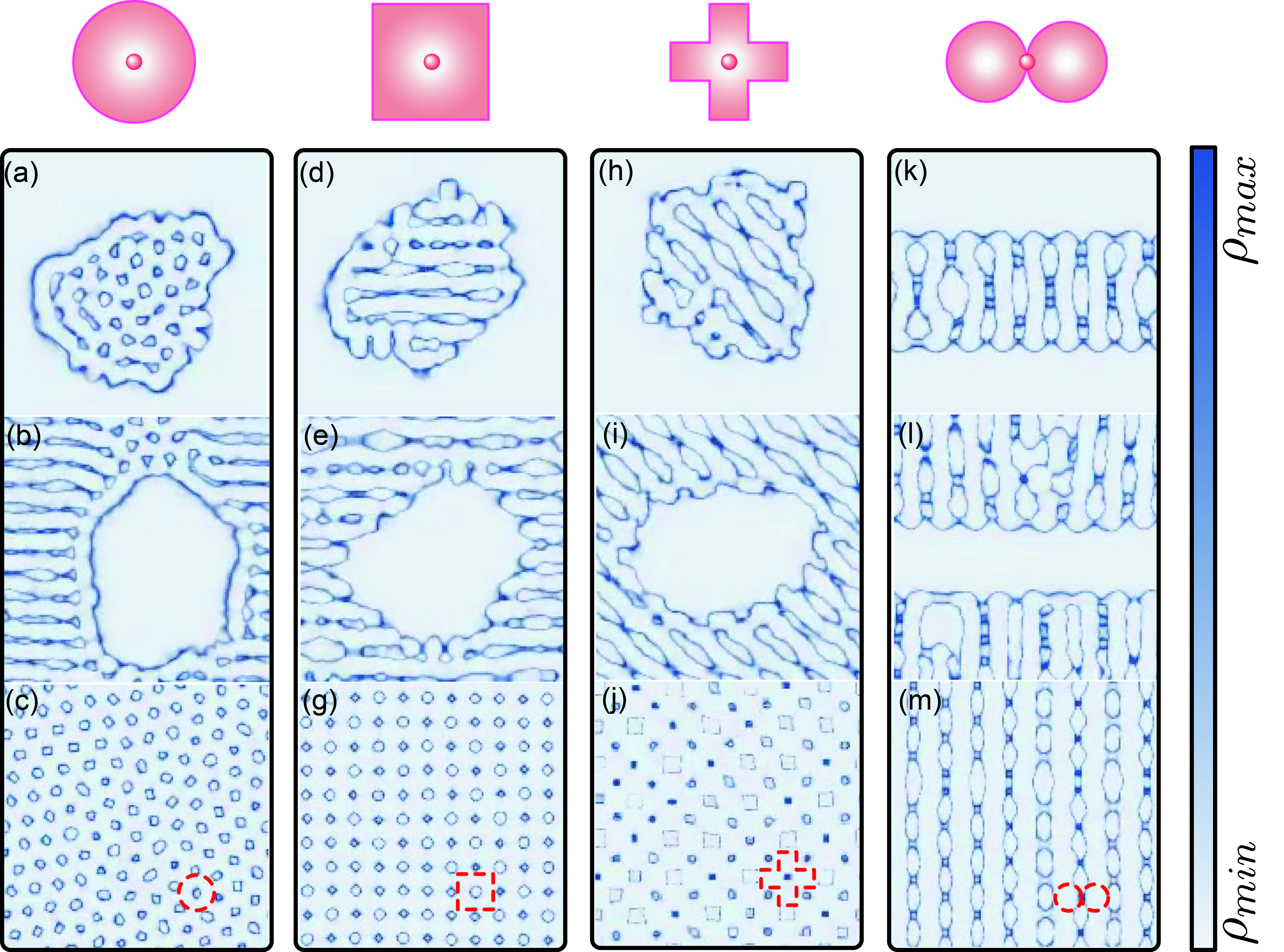}
    \caption{\textbf{Continuum-field simulations with different perception kernels.}
    Representative steady-state density fields obtained from the deterministic continuum model for several finite-range perception geometries. The icons above each column indicate the corresponding perception kernel: circular, square, cross-shaped, and double-circle sensing regions. For each geometry, finite-range perception generates internally patterned dense domains at intermediate densities and system-spanning microphase-separated states at higher densities. The continuum theory qualitatively reproduces the particle-level morphologies, including hybrid phase separation, ordered bubble structures, and geometry-dependent anisotropic pattern selection.}
\label{fig:SFieldKernels}
\end{figure}
As shown in Fig.~\ref{fig:SFieldKernels}, the continuum theory qualitatively reproduces the homogeneous, hybrid phase-separated, and microphase-separated morphologies observed in the particle model. In particular, it captures both the emergence of internally patterned dense domains and the transition to system-spanning microphase order. This agreement supports Eq.~\eqref{eq:continuum_main} as a minimal field theory for perception-induced organization.

\subsection{Simulation protocol for one-dimensional perception}

Now we introduce the protocol of the simulation shown in Fig.~4(a) of the main text. We simulate the one-dimensional version of the continuum equation on a periodic domain of length $L=150$. The domain is discretized into $N=450$ lattice sites, corresponding to a spatial resolution $h=1/3$. The density field evolves according to the same conserved dynamics used in two dimensions and is integrated by a finite-difference scheme with time step $\Delta t=10^{-3}$. We use a stabilizing hyperdiffusion coefficient $\kappa=0.01$ and a perception radius $r_0=15$. The system is initialized near a homogeneous state with a single weak sinusoidal perturbation,
\begin{equation}
    \rho(x,0)
    =
    \rho_0
    +
    A\cos\left(\frac{2\pi n}{L}x\right),
\end{equation}
where $\rho_0=1.05$, $A=0.01$, and $n=8$. This choice selectively excites the fundamental patterning mode in the MPS regime, allowing its linear growth and subsequent nonlinear hollowing into a one-dimensional bubbling state to be resolved.

% ========================================================
\section{Linear stability analysis}

This section shows detailed derivation and analysis of the linear stability of the continuum equation. To this end, we introduce the following convention for the Fourier series of the density field defined on a domain of size $L \times L$:
\begin{equation}
\rho(\mathbf{r},t)=\sum_{n,m}\exp{i k_n r_1+i k_m r_2}\rho_{n,m}(t) \equiv\sum_{\mathbf{n}} \exp{i \mathbf{k}_{\mathbf{n}}\cdot \mathbf{r}} \rho_{\mathbf{n}}(t)\ ,
\end{equation}
where we use the vector index $\mathbf{n}=[n,m ]^\text{T}$, the wavevector $\mathbf{k}_{\mathbf{n}}=(2\pi/L)[n,m ]^\text{T}$, and the Fourier amplitude $\rho_{\mathbf{n}}=\rho_{[n,m]}$ to simplify the notation. The zero mode $\mathbf{n}=\mathbf{0}$ corresponds to the conserved homogeneous background density, such that $\rho_{\mathbf{0}} = \rho_0$. The non-local perception density is defined via a convolution $\bar{\rho}(\mathbf{r}) = \int \dint {\mathbf{r}'} K(\mathbf{r}-\mathbf{r}')\rho(\mathbf{r}')$. In reciprocal space, this convolution simplifies to a direct product: $\bar{\rho}_{\mathbf{n}} = \widehat K(\mathbf{k}_{\mathbf{n}})\rho_{\mathbf{n}}$, where $\widehat K(\mathbf{k})$ is the Fourier transform of the perception kernel. For a uniform sensing region $\Omega$ of area $S_\Omega$, the normalized kernel is:
\begin{equation}\widehat K(\mathbf{k}_{\mathbf{n}}) = \frac{1}{S_\Omega} \int_\Omega \exp{-i \mathbf{k}_{\mathbf{n}}\cdot\mathbf{r}'} \dint{\mathbf{r}}'\ .
\end{equation}
To obtain the time evolution of the Fourier modes, we substitute the Fourier series into the continuum dynamics $\partial_t \rho = \nabla^2 [D(\bar{\rho})\rho] - \kappa \nabla^4 \rho$. Instead of assuming a specific functional form for the mobility, we perform a general Taylor expansion of $D(\bar{\rho})$ around the homogeneous background density $\rho_0$:
\begin{equation}D(\bar{\rho}) = D(\rho_0) + D'(\rho_0)\delta\bar{\rho} + \mathcal{O}(\delta\bar{\rho}^2)\ ,
\end{equation}
where $\delta\bar{\rho} = \bar{\rho} - \rho_0$. Substituting this into the chemical potential $\mu_{\rm P} = D(\bar{\rho})\rho$, applying the Laplacian operator $\nabla^2 \to -k_{\mathbf{n}}^2$ and setting $\rho_{\mathbf{n}}(t) \sim \exp{\sigma(\mathbf{k}_{\mathbf{n}}) t}$, we obtain the dispersion relation
\begin{equation}
    \label{eq:SDispRela}
     \sigma(\mathbf k)
    =
    -k^2
    \left[
        D(\rho_0)
        +
        \rho_0D'(\rho_0)\widehat K(\mathbf k)
        +
        \kappa k^2
    \right],
\end{equation}
presented in the main text. 

%%%%
The function $\widehat K(\mathbf k)$ acts as a wavevector-dependent spectral weight, encoding how the perception geometry selects the Fourier modes that enter the instability. For standard sensing domains, $\widehat K(\mathbf k)$ can be obtained analytically. For a circular perception region of radius $r_0$, the normalized Fourier kernel for nonzero modes is
\begin{equation}
\label{eq:transform_circular}
\widehat K(\mathbf k_{\mathbf n})
=
\frac{1}{\pi r_0^2}
\int_0^{2\pi} d\theta
\int_0^{r_0} dr\, r\,
e^{-i k r\cos\theta}
=
\frac{2J_1(k r_0)}{k r_0},
\end{equation}
where $k=|\mathbf k_{\mathbf n}|$ and $J_1$ is the Bessel function of the first kind. We also use a quasi-one-dimensional perception domain, a rectangle of size $2r_0\times L$, to isolate the mechanism of cluster formation. Its Fourier kernel is
\begin{equation}
\label{eq:quasi_one_dimensional}
\widehat K(\mathbf k_{\mathbf n})
=
\frac{1}{2r_0L}
\int_{-r_0}^{r_0} dx
\int_{-L/2}^{L/2} dy\,
e^{-(i k_n x+i k_m y)}
=
\frac{\sin(k_n r_0)}{k_n r_0}\delta_{m,0}.
\end{equation}
The factor $\delta_{m,0}$ reflects translational invariance along the transverse direction, so only density modulations along $x$ are coupled by this quasi-one-dimensional sensing kernel.

\begin{figure}
    \centering
    \includegraphics[width=1\linewidth]{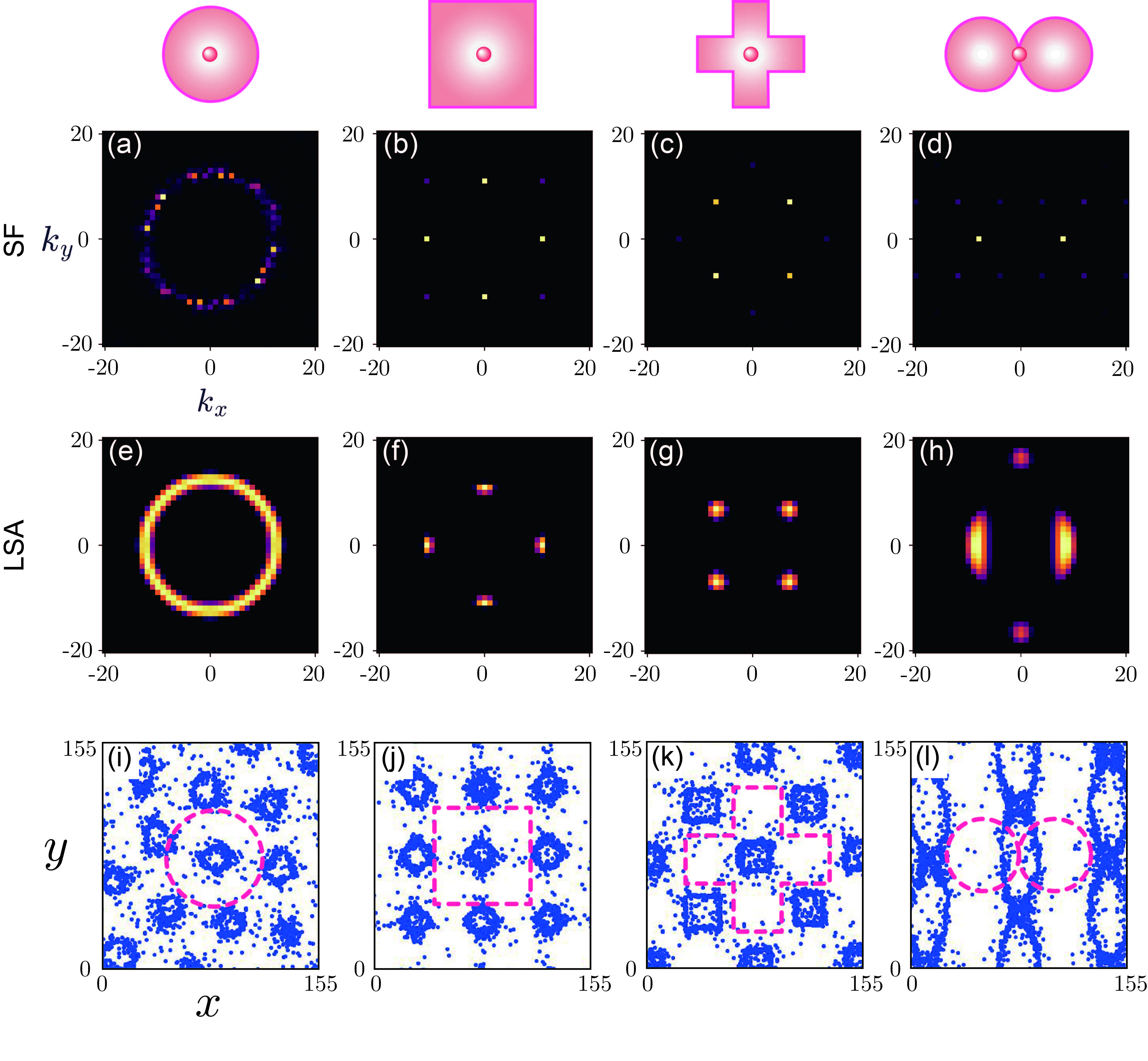}
    \caption{\textbf{Structure factors and linear-stability prediction for different perception kernels.}
    Comparison between particle simulations and linear stability analysis for circular, square, cross-shaped, and double-circle perception geometries. The first row shows the structure factor $S(\mathbf{k})$ measured from particle configurations, where bright peaks indicate the dominant density modes selected by the steady-state pattern. The second row shows the corresponding linearly unstable modes predicted from the dispersion relation of the continuum theory. The third row presents representative particle configurations, with red dashed curves marking the shape and orientation of the perception zone. Across different sensing geometries, the dominant peaks in $S(\mathbf{k})$ agree with the unstable modes predicted by linear stability analysis, demonstrating that the perception kernel selects both the characteristic wavelength and the symmetry of the emergent microstructure.}
\label{fig:SSFLSA}
\end{figure}

To examine the effect of broken rotational symmetry, we also consider anisotropic perception domains. For a square perception region of side length $a$ centered at the origin, the normalized kernel factorizes as
\begin{equation}
\label{eq:kernel_square}
\widehat K(\mathbf k)
=
\operatorname{sinc}\left(\frac{k_xa}{2}\right)
\operatorname{sinc}\left(\frac{k_ya}{2}\right)
=
\frac{4}{k_xk_ya^2}
\sin\left(\frac{k_xa}{2}\right)
\sin\left(\frac{k_ya}{2}\right),
\end{equation}
where $\operatorname{sinc}(z)=\sin z/z$.

For a cross-shaped perception region constructed from five identical squares of side length $\ell/3$, with total bounding size $\ell$, the kernel follows by superposing the horizontal and vertical arms and subtracting their overlap:
\begin{align}
\label{eq:kernel_plus}
\widehat K(\mathbf k)
&=
\frac{1}{5}
\left[
3\operatorname{sinc}\left(\frac{k_x\ell}{2}\right)
 \operatorname{sinc}\left(\frac{k_y\ell}{6}\right)
+
3\operatorname{sinc}\left(\frac{k_x\ell}{6}\right)
 \operatorname{sinc}\left(\frac{k_y\ell}{2}\right)
-
\operatorname{sinc}\left(\frac{k_x\ell}{6}\right)
 \operatorname{sinc}\left(\frac{k_y\ell}{6}\right)
\right]
\nonumber \\
&=
\frac{36}{5\ell^2 k_x k_y}
\left[
\sin\left(\frac{k_x\ell}{2}\right)
\sin\left(\frac{k_y\ell}{6}\right)
+
\sin\left(\frac{k_x\ell}{6}\right)
\sin\left(\frac{k_y\ell}{2}\right)
-
\sin\left(\frac{k_x\ell}{6}\right)
\sin\left(\frac{k_y\ell}{6}\right)
\right].
\end{align}

Finally, for a double-circle perception region consisting of two tangent disks of radius $r_0$ centered at $(\pm r_0,0)$, the Fourier shift theorem gives
\begin{equation}
\label{eq:kernel_double_circle}
\widehat K(\mathbf k)
=
\cos(k_x r_0)\,
\frac{2J_1(k r_0)}{k r_0},
\end{equation}
with $k=\sqrt{k_x^2+k_y^2}$. All expressions are understood in their continuous limits when a denominator vanishes, so that $\widehat K(\mathbf 0)=1$ for each normalized perception kernel.

This mode selection is directly reflected in the simulated structures. The unstable modes predicted by the linear dispersion relation in Eq.~(\ref{eq:SDispRela}) correspond to the dominant peaks in the structure factor of the ordered states. As shown by comparing the structure factors from particle simulations with the dispersion relations (Fig. \ref{fig:SSFLSA}), the peaks of $S(\mathbf k)$ align with the fastest-growing modes of $\sigma(\mathbf k)$, confirming that the emergent spatial order is selected by the perception-mediated linear instability.

% ======================================================
\section{Kinematic Origin of the Bubbling Phase}
\label{sec:kinematic_origin}

In this section, we analyze the local mechanism by which a growing density peak hollows into a bubble. Using a single-mode ansatz, we show that nonlocal perception can invert the curvature of the perception-mediated chemical potential at the peak center once the mode amplitude exceeds a critical value. This inversion reverses the local conserved flux and drives material from the peak center toward its shoulders, thereby exciting higher harmonics that produce the hollow core. Here, we focus on the case $\rho_0>\rho_c$.

The one-dimensional continuum dynamics is governed by 
\begin{equation}
    \partial_t \rho = \partial_x^2 \mu_{\rm P} - \kappa \partial_x^4 \rho \ ,
\end{equation}
where the nonlocal chemical potential is $\mu_{\rm P}(x)=D[\bar{\rho}(x)]\rho(x)$. In this analysis we neglect the small bare diffusivity $\chi$ since it contributes a local term to chemical potential but does not change the curvature-inversion mechanism as presented later. Therefore, we use $D(\bar{\rho})=D_0(\bar{\rho}/\rho_c-1)^2$. To isolate the nonlinear step, we approximate the density by the fundamental mode 
\begin{equation}
    \rho(x)=\rho_0+A\cos(k_*x), 
\end{equation}
where $A>0$ and $x=0$ is the center of a density peak. The corresponding perceived density is 
\begin{equation}
\bar{\rho}(x)=\rho_0+A\widehat K_*\cos(k_*x),
\end{equation}
with $\widehat K_*\equiv\widehat K(k_*)$. Defining the perceived-density bias 
\begin{equation}
f(x)=\bar{\rho}(x)-\rho_c=(\rho_0-\rho_c)+A\widehat K_*\cos(k_*x),
\end{equation}
the chemical potential becomes $\mu_{\rm P}=(D_0/\rho_c^2)f(x)^2\rho(x)$.

We now evaluate the curvature of $\mu_{\rm P}$ at the peak center. At $x=0$, the relevant quantities are $\rho(0)=\rho_0+A$, $\rho'(0)=0$, $\rho''(0)=-Ak_*^2$, $f(0)=\rho_0-\rho_c+A\widehat K_*$, $f'(0)=0$, and $f''(0)=-A\widehat K_*k_*^2$. Applying the product rule twice to $\mu_{\rm P}(x)=(D_0/\rho_c^2)f(x)^2\rho(x)$ gives
\begin{equation}
\mu_{\rm P}''(x)=(D_0/\rho_c^2)\bigg(2[f'(x)^2+f(x)f''(x)]\rho(x)+4f(x)f'(x)\rho'(x)+f(x)^2\rho''(x)\bigg).
\end{equation}
Since the first derivatives vanish at the peak center, this reduces to 
\begin{equation}
\mu_{\rm P}''(0)=-(D_0/\rho_c^2)Ak_*^2f(0)[2\rho(0)\widehat K_*+f(0)].
\end{equation}

We focus on the dense regime $\rho_0>\rho_c$, where the finite-$k_*$ instability occurs on a negative lobe of the perception kernel, $\widehat K_*<0$. The bracketed factor can be written as 
\begin{equation}
    2\rho(0)\widehat K_*+f(0)=(\rho_0-\rho_c+2\rho_0\widehat K_*)+3A\widehat K_*.
\end{equation}
The first term is proportional to the linear chemical-potential response of the fundamental mode and is negative for the unstable mode considered here, while the nonlinear contribution $3A\widehat K_*$ is also negative because $A>0$ and $\widehat K_*<0$. Thus the bracket remains negative, and the sign of $\mu_{\rm P}''(0)$ is controlled by the perceived-density bias $f(0)$.

This gives a two-stage picture of bubbling. For small amplitudes, $f(0)>0$, so $\mu_{\rm P}''(0)>0$: the peak center is a local minimum of the chemical potential, and the conserved current drives material into the density maximum, amplifying it. As the fundamental mode grows, the negative kernel value lowers the perceived density at the peak center. When $A$ exceeds the critical value $A_c=(\rho_0-\rho_c)/|\widehat K_*|$, the bias $f(0)$ changes sign. The curvature then becomes negative, $\mu_{\rm P}''(0)<0$, so the peak center becomes a local maximum of the chemical potential. The conserved current reverses and transports material outward from the center toward the shoulders. A pure cosine mode cannot accommodate this redistribution, so higher spatial harmonics are generated; in real space, the compact peak flattens and hollows into a bubble.

\bibliographystyle{apsrev4-1}
\bibliography{QS_Brownian}
% ========================================================
%\begin{thebibliography}{9}

%\bibitem{ref1}
%A. Author and B. Author,
%\textit{Journal Name} \textbf{12}, 345 (2024).

%\end{thebibliography}

%% file: macros.tex
\usepackage{dsfont}

\newcommand{\canetset}[1]{{\mathchoice {\hbox{$\sf\textstyle #1\kern-0.4em #1$}}
{\hbox{$\sf\textstyle #1\kern-0.4em #1$}}
{\hbox{$\sf\scriptstyle #1\kern-0.3em #1$}}
{\hbox{$\sf\scriptscriptstyle #1\kern-0.2em #1$}}}}

\def\nbZ{{\mathchoice {\hbox{$\sf\textstyle Z\kern-0.4em Z$}}
{\hbox{$\sf\textstyle Z\kern-0.4em Z$}}
{\hbox{$\sf\scriptstyle Z\kern-0.3em Z$}}
{\hbox{$\sf\scriptscriptstyle Z\kern-0.2em Z$}}}}
\usepackage{bm}

\newlength \standardfigwidth
\newcounter{exercise}
{\addtocounter{exercise}{1}\begin{center}\begin{minipage}{0.8\linewidth}\textbf{Exercise
\arabic{exercise}:}\begin{itshape}}% begin code
{\end{itshape}\end{minipage}\end{center}}%                    end code

\makeatletter
\newcommand{\creat}[3][]{\@ifempty{#1}{#2^{\dagger}}{\left(#2^{\dagger}\right)^{#1}}\@ifempty{#3}{}{\!(#3)}}

\newcommand{\creatDoi}[3][]{\@ifempty{#1}{\tilde{#2}}{\left(\tilde{#2}\right)^{#1}}\@ifempty{#3}{}{(#3)}}

\newcommand{\annih}[3][]{#2\@ifempty{#1}{}{^{#1}}\@ifempty{#3}{}{(#3)}}

\makeatother

% ...
\newlength{\bibmarkkeyAleft}

\newlength{\bibmarkkeyBleft}

\newlength{\bibmarkkeyCleft}

\newlength{\bibmarkkeyDleft}